\documentclass[letterpaper]{article} 
\usepackage{aaai25}  
\usepackage{times}  
\usepackage{helvet}  
\usepackage{courier}  
\usepackage[hyphens]{url}  
\usepackage{graphicx} 
\usepackage{natbib}  
\usepackage{caption} 
\usepackage{algorithm}
\usepackage{algorithmic}
\usepackage{multirow}
\usepackage{multicol}
\usepackage{amssymb}
\usepackage{subcaption}
\usepackage{makecell}
\usepackage{amsmath}

\usepackage{newfloat}
\usepackage{listings}
\DeclareCaptionStyle{ruled}{labelfont=normalfont,labelsep=colon,strut=off} 
\floatstyle{ruled}
\newfloat{listing}{tb}{lst}{}
\floatname{listing}{Listing}
\title{BeniFul: Backdoor Defense via Middle Feature Analysis \\ for Deep Neural Networks}
\author {
    Xinfu Li\textsuperscript{\rm 1},
    Junying Zhang\textsuperscript{\rm 1},
    Xindi Ma\textsuperscript{\rm 1}
}
\affiliations {
    \textsuperscript{\rm 1}Xidian University\\
    xinfuli@stu.xidian.edu.cn, \{jyzhang1990, xdma1989\}@gmail.com
}

\begin{document}

\maketitle

\begin{abstract}
Backdoor defenses have recently become important in resisting backdoor attacks in deep neural networks (DNNs), where attackers implant backdoors into the DNN model by injecting backdoor samples into the training dataset. Although there are many defense methods to achieve backdoor detection for DNN inputs and backdoor elimination for DNN models, they still have not presented a clear explanation of the relationship between these two missions. In this paper, we use the features from the middle layer of the DNN model to analyze the difference between backdoor and benign samples and propose Backdoor Consistency, which indicates that at least one backdoor exists in the DNN model if the backdoor trigger is detected exactly on input. By analyzing the middle features, we design an effective and comprehensive backdoor defense method named BeniFul, which consists of two parts: a gray-box backdoor input detection and a white-box backdoor elimination. Specifically, we use the reconstruction distance from the Variational Auto-Encoder and model inference results to implement backdoor input detection and a feature distance loss to achieve backdoor elimination. Experimental results on CIFAR-10 and Tiny ImageNet against five state-of-the-art attacks demonstrate that our BeniFul exhibits a great defense capability in backdoor input detection and backdoor elimination.
\end{abstract}

\section{Introduction}
Deep learning, as a typical branch of machine learning, has been widely applied in our lives, such as autonomous driving~\cite{Hu2024DALDet}, medical care~\cite{Huang2024Sparse}, education~\cite{Nancye2024AI}, etc. However, deep neural network (DNN) models are shown to be vulnerable to backdoor attacks~\cite{Wenger2021Backdoor}, which attacks include two steps: backdoor implantation and backdoor triggering. Specifically, backdoor attackers could implant a backdoor into a targeted DNN model by injecting a few backdoor samples into the training set. During the training process, the DNN model will learn a strong correlation between the backdoor trigger and the target label, which does not even affect the model's performance.

According to the impact of backdoor triggers on samples, existing backdoor attack methods can be divided into three categories: patch-based triggers which are patterns patched on input samples, such as Blend~\cite{Chen2017Targeted}, BadNets~\cite{Gu2019BadNets}, and PatchBackdoor~\cite{Yuan2023PatchBackdoor}, transform-based triggers which are invisible transformation on benign samples, such as WaNet~\cite{Nguyen2021WaNet}, BATT~\cite{Xu2023BATT} and DT-IBA~\cite{Sun2024Invisible}, and physical triggers which are elements of our physical world, such as PhysicalBA~\cite{Wenger2021PhyasicalBA}, Kaleidoscope~\cite{gong2023kaleidoscope} and SRA~\cite{qi2022towards}. In the model inference stage, the inference result will be manipulated by attackers through the backdoor trigger, which will cause serious consequences for model users.

To resist backdoor attacks, many defense schemes have been proposed, such as separating backdoor samples from the training set~\cite{Zhou2024DataElixir, Gao2023Backdoor}, training a clean DNN model on backdoor dataset~\cite{Li2021ABL, Zhang2023CBD}, backdoor input detection~\cite{Gao2019STRIP, Guo2023SCALE}, and eliminating the backdoor in DNN model~\cite{Li2021NAD, Liu2022BAERASER}. These schemes are around the lifecycle of the DNN model. Before training, the main defense is sweeping the backdoor samples away from the train set. During the training stage, resisting backdoor implantation is used to achieve backdoor defense. After training or in the inference stage, defense methods include backdoor model detection, backdoor input detection, and backdoor elimination of models. This paper focuses on backdoor defense in a trained model, specifically including backdoor input detection and model backdoor elimination.

As a typical backdoor input detection method, cycling inferring the test samples multiple times during model runtime can achieve good detection, but it will also seriously reduce the model efficiency, such as STRIP~\cite{Gao2019STRIP}, SCALE-UP~\cite{Guo2023SCALE} and TeCo~\cite{Liu2023TeCo}, whose detection manner is referred to as Inference Consistency in this paper. Compared to using this Inference Consistency, defenders can fully utilize known clean datasets to achieve more efficient backdoor input detection in runtime, such as less inferring times.
Especially when the defender is the trainer or one of the trainers, there is high credibility for his own dataset, and the defender could make full use of this dataset for better backdoor detection and elimination.

Unlike previous methods, we achieve the backdoor defense by analyzing the features from the DNN model's middle layer, where the features are referred to as the middle features in this paper. Our detection and elimination methods are based on two truths, which are also stated in~\cite{Fu2023DifferentialAnalysis}. One is that the middle features extracted by the backdoor model are different between the normal samples and the backdoor samples. The other is that the middle features extracted from the backdoor sample are different between the normal model and the backdoor model. We validate these differences by comparing the benign middle feature map with different backdoor middle feature maps, which are verified in Section~\ref{sec: backdoor consistency}, and obtain the fact that different backdoor trigger types in inputs cannot trigger each other backdoors in the model, which is called Backdoor Consistency in this paper. Backdoor Consistency means that when we detect the existence of backdoor inputs reliably, at least one type of backdoor exists in the DNN model. Based on the Backdoor Consistency, we connect the relationship between the backdoor input detection and the backdoor elimination.

In this paper, we propose a backdoor defense method based on the analysis of middle features, named BeniFul, incliuding gray-box backdoor input detention and white-box backdoor elimination. For backdoor input detection, we train a Variational Auto-Encoder (VAE) model~\cite{Diederik2014VAE} by the middle features from the middle layer of the target model. And we determine whether a test input is a backdoor sample by jointly analyzing the VAE reconstruction results and the inference results of the target model. For the backdoor elimination, we define a loss function that makes the middle features of the eliminated model far away from the original backdoor model. We add this loss function to the model's task loss function to train the target model, eradicate the backdoor in the target model, and maintain the model's accuracy as much as possible. The main contributions are summarized as follows.


\begin{itemize}
    \item To achieve a more comprehensive backdoor defense, we propose Backdoor Consistency which provides a theoretical basis to integrate different kinds of defense methods. We utilize this Backdoor Consistency to link the backdoor input detection and backdoor elimination in our defense method.


    \item Based on the difference in middle features between benign and backdoor inputs, we propose a gray-box backdoor input detection method, named BeniFul-BID, which could detect the backdoor input with only once model inference. Then, by maximizing the knowledge difference between the eliminated model and the original backdoor model, a backdoor elimination method, named BeniFul-BE, is also designed to repair the backdoored model.
    \item We conduct comprehensive experiments to evaluate our detection and elimination method under five backdoor attacks, which achieves effective detection with about $0.953$ average AUROC score and substantive elimination with about $0.967$ average ASR decline and only $0.028$ average ACC loss over Tiny ImageNet.
\end{itemize}



\section{Preliminaries}



\subsection{Backdoor Attack}
This paper focuses on dirty-label backdoor attacks in which attackers can use data poisoning attacks or directly participate in model training to implement backdoors into DNN models.
For an example of image classification tasks, the attacker could generate $N$ backdoor image samples $\{x^t_i\}^N_{i=1}$ by adding a patch $t$ on them, $x^t_i = m \odot x_i + (1-m)\odot t$, transforming them, $x^t_i = transform(x_i; t, m)$, or selecting physical world elements as trigger $t$. Then, the attacker makes $y^t_i$ as the target backdoor label for $x^t_i$ and puts $(x^t_i, y^t_i)$ into the training set $D_{train}$ to implant the backdoor into the target DNN model $M$:
$$\theta^b = \mathop{\arg\min}\limits_{\theta} \frac{1}{N_0 + N}\sum_{i=1}^{N_0 + N} \mathcal{L}(M(\theta; \hat{x}_i), \hat{y}_i) $$
where $(\hat{x}_i, \hat{y}_i) \in D_{train} \cup \{x^t_i,y^t_i\}^N_{i=1}$ and, $N_0$ represents the size of $D_{train}$. The backdoor model exhibits good classification for benign samples and a high attack success for backdoor inputs. So in the model inference stage, attackers can modify an input sample $x_\diamond$, implant a backdoor trigger $t$ to make it $x^t_\diamond$, and cause the model to make a backdoor inference at least with an attack success ratio of $\eta$:
$$ \mathcal{P}(M(x^t_\diamond; \theta^b) = y^t) \geq \eta. $$

\subsection{Variational Auto-Encoder}\label{sec: vae}
Different from auto-encoders (AE), the Variational Auto-Encoder (VAE) adds constraints on the latent feature space and makes this space continuous.
For the sample $x_i$ in the dataset $D$, the encoder model $q_{\phi}(z|x_i)$ extracts the latent variable $z$ from input $x_i$, where $\phi$ are the parameters of the encoder model $q$.
The decoder model $p_\theta(x_i|z)$ produces a distribution over the possible corresponding values of $x_i$ according to $z$ where $\theta$ are the parameters of the decoder model $p$.
The optimized variational lower bound for decoding model parameters is:
$$ \mathcal{L}(\theta,\phi;x_i) = -\mathrm{D}_{KL} + \mathrm{E}_{q_{\phi}(z | x_i)}[\log p_\theta(x_i | z)]$$
where $-\mathrm{D}_{KL}$ is the Kullback-Leibler (KL) divergence loss  from $q_{\phi}(z | x_i)$ to $p_{\theta}(z)$ and $p_\theta(z)$ represents the prior probability of the potential spatial variable $z$.
When the sample $x_i$ is an image and is fed into the neural network encoder, the output is the vector of mean $\mu$ and variance $\sigma$ for potential features. The latent variable $z$ could be obtained through the reparameterization function $z = \mu + \sigma \odot \epsilon$, where $\epsilon$ corresponds to the normal distribution and $\odot$ represents element-wise
product. Hence, the logarithmic form of the variational approximation posterior probability obtained is
 $\log q_\phi(z|x_i) = \log \mathcal{N}(z;\mu^i,(\sigma^i)^2I)$. The KL loss in the optimization function turns into:
$$ -\mathrm{D}_{KL} = \frac{1}{2} \sum_{j=1}^{J} (1 + \log((\sigma_j^i)^2) - (\mu_j^i)^2 - (\sigma_j^i)^2) $$
where $J$ is the dimension of the laten variable $z^{(i,l)}$, $z^{(i,l)} = \mu^i + \sigma^i \odot \epsilon^{l}$, and $\epsilon^{l} \sim \mathcal{N}(0,1)$. Whether the optimization function is for a probability model or an image reconstruction model, it is not difficult to observe that it not only includes reconstruction loss for the input sample but also regularizes the latent variables from the encoder.
\subsection{Threat Model.}
In the training stage, we assume that the attacker has full access to the training dataset and white box access to the target model. Attackers can apply any backdoor attack method to attack DNN models, and the backdoor trigger pattern can be of any shape, position, and size. In the inference stage, the attacker could manipulate the input and trigger the backdoor in the target model. We also assume
that the backdoor model exhibits good performance for normal samples and a high attack success ratio for backdoor inputs.

Backdoor defense, in this paper, refers to two parts: backdoor detection and backdoor elimination, and the defenders include backdoor detector and backdoor eliminator:
a backdoor detector performs backdoor detection on the unknown model's inputs and a backdoor eliminator performs backdoor elimination for the backdoor model after detection.
In our scheme, the detector needs to access a small amount of clean dataset and gray-box access to the model to detect whether the input sample is a backdoor sample or benign. This gray-box access only obtains the inference label and the middle feature outputs from the middle layer of the detected model.
In the process of eliminating backdoors, the eliminator needs to use a small portion of the clean dataset and white-box access to the target DNN model. This access can obtain the model's structure, parameters, and gradients to fine-tune the backdoor model.

\section{Details of Our BeniFul}

\subsection{Backdoor View from the Middle Feature}
\begin{figure}[!htbp]
    \centering
    \includegraphics[width=0.47\textwidth]{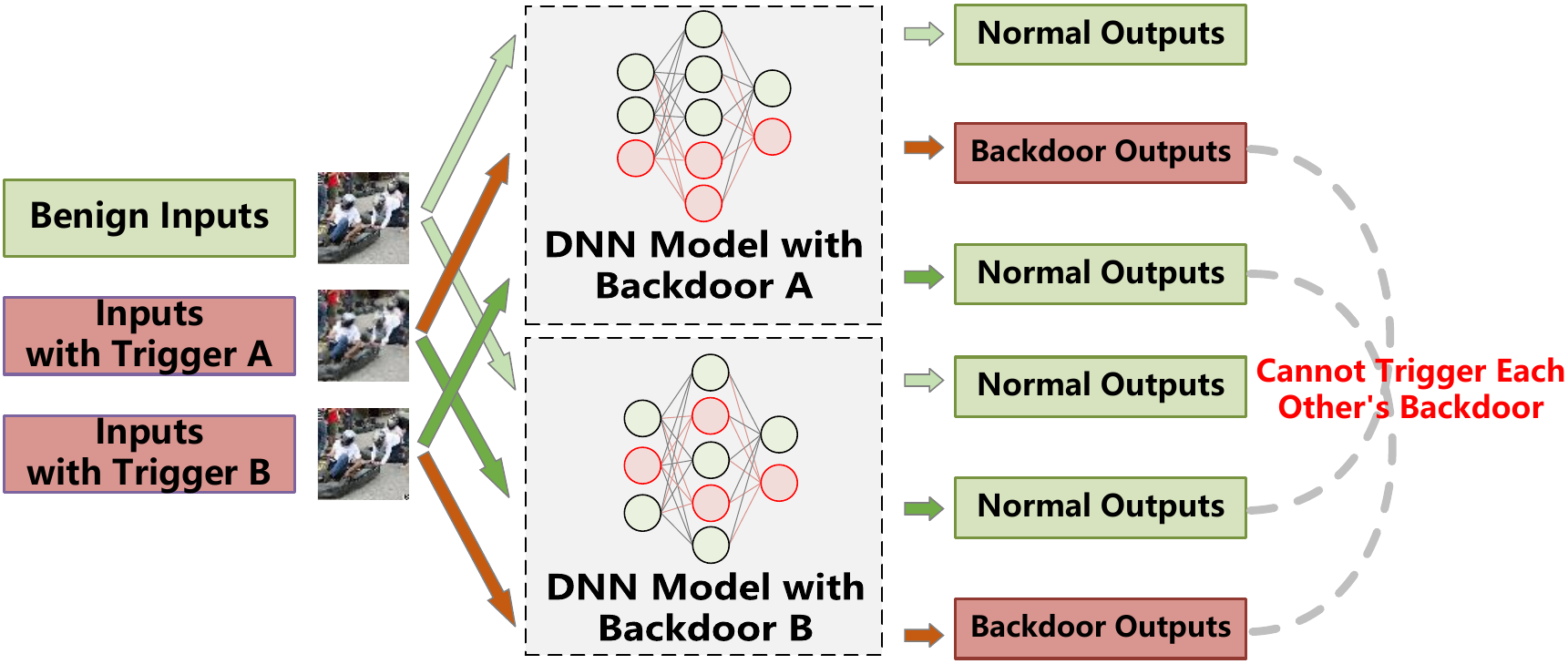}
    \caption{Illustration of Backdoor Consistency.}
    \label{fig: backdoor consistency}
\end{figure}

It is obvious that the backdoor implanted in a DNN model has a consistency corresponding to the backdoor trigger on inputs, where the consistency is referred to as Backdoor Consistency in this paper. That is, two unrelated backdoor triggers cannot activate each other backdoor in a DNN model, as shown in Figure~\ref{fig: backdoor consistency}.
The process of feature extraction in the model with the backdoor $A$ will not be misled by the backdoor trigger $B$.
For example, in the backdoor model under the WaNet attack, the middle features extracted from the BadNet backdoor input are similar to the middle features of benign samples, as shown in Figure~\ref{fig: middle feature}.

It can be seen from the perspective of middle features that, as detectors, we could distinguish the distribution between benign features and backdoor features to achieve backdoor detection at the input level. Based on the Backdoor Consistency, we can determine whether there is a backdoor in the DNN model correspondingly. For the already detected backdoor model, as eliminators, we could unify the distribution of backdoor features to the distribution of benign features to achieve backdoor elimination.

\subsection{BeniFul - Backdoor Input Detection}
\begin{figure}[!htbp]
    \centering
    \includegraphics[width=0.47\textwidth]{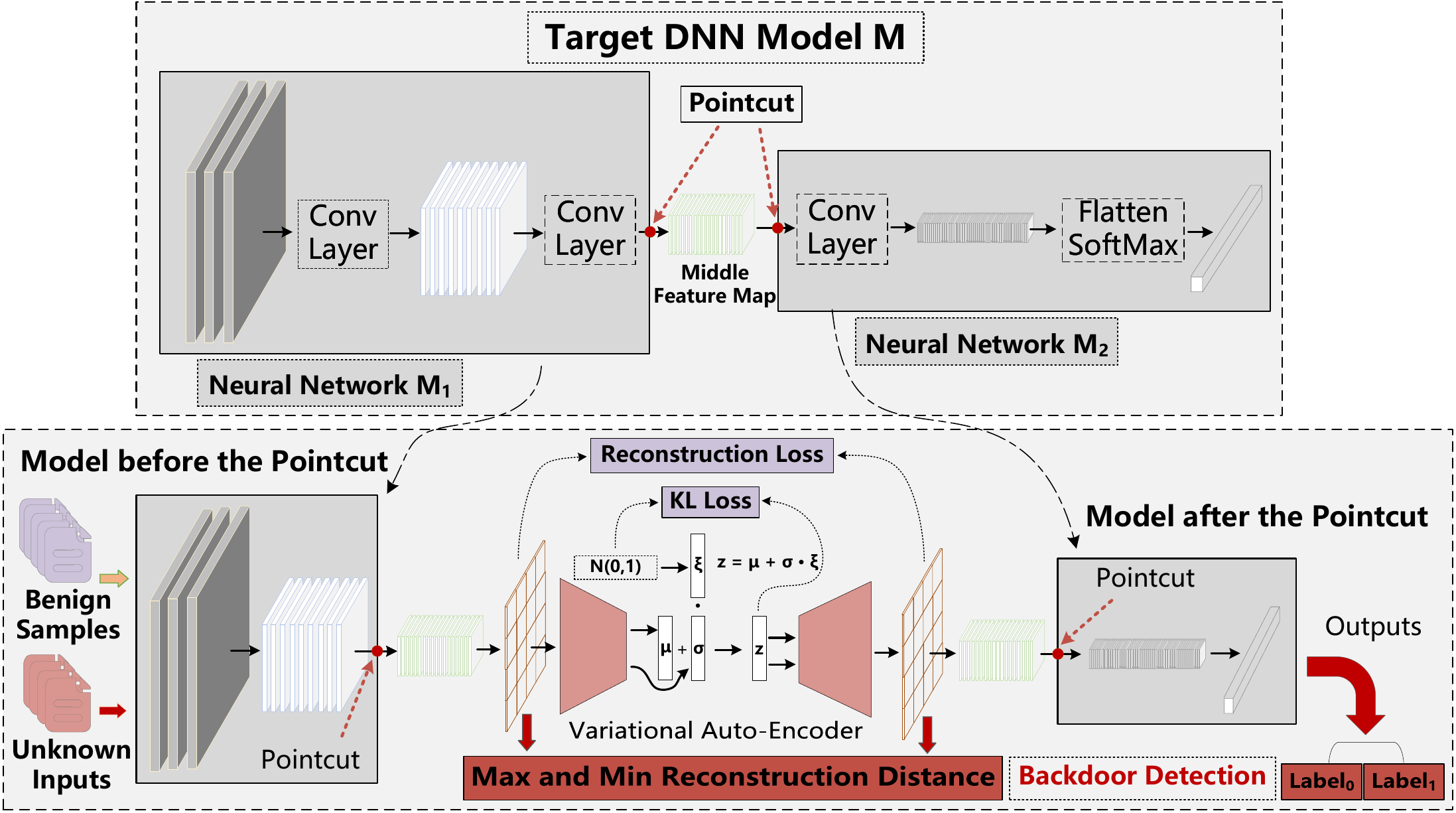}
     \caption{Backdoor Input Detection.} \label{fig: detection architecture}
\end{figure}

Motivated by the truth of the difference in middle features between backdoor inputs and benign inputs, we use a VAE model to reconstruct benign intermediate features. Then, we further detect the backdoor input through the difference in the target model inference results and the distribution of the VAE reconstruction distance.
As shown in Figure~\ref{fig: detection architecture}, for the DNN model to be detected, we only need to access the intermediate features from the ``pointcut'' and the inference result output from the model.

\subsubsection{Detection Framework.} For a DNN model $M$ to be detected, we refer to the position where we could obtain the middle features as the ``pointcut'', and formally divide the model into two parts by the pointcut: the model before pointcut is denoted as $M_1$ and the model after pointcut is denoted as $M_2$.
For an input sample $x$, the inference process of the target model is divided into two parts, $m = M_1 (x)$ and $y=M_2 (m)$, where $m$ are the middle features output from the pointcut. We reshape $m$ into a three-dimensional feature map $\bar{m} = reshape(m)$ sized like $1\times w \times h$. Then we could build a VAE training set $\{\bar{m}_i\}^{N_1}_{i = 1}$ and a test set $\{\bar{m}_i\}^{N_2}_{i = 1}$ from a part of the clean train set. It should be noted that our scheme is completely black-box access to the $M_1$ and $M_2$ models because we do not need to know the parameters, gradients, and model architecture of these two models. Moreover, for the subsequent training VAE and input detection, we do not even need to know the truth labels of the samples or inputs.

\subsubsection{Training of VAE.} To achieve backdoor detection, the $\{\bar{m}_i\}^{N_1}_{i = 1}$ is extracted from benign samples and used to train the VAE model.
In our scheme, the training loss function of VAE also includes two parts: the KL loss and the reconstruction loss. Therefore, our VAE training loss function is:
\begin{equation*}
    \begin{split}
        \mathcal{L}(\theta_{vae}; \bar{m}) = & MSE(\bar{m}, VAE(\bar{m}; \theta_{vae}))\\
        & + \alpha \cdot \sum_{j=1}^{J} ((\mu_j)^2 + (\sigma_j)^2) - \log((\sigma_j)^2) - 1 ,
    \end{split}
\end{equation*}
where $\theta_{vae}$ is the weight of the VAE model, $J$ is the dimension of the latent variable $z$, $z = \mu + \sigma \odot \epsilon$, MSE represents Mean-Square Error,  $\epsilon \sim \mathcal{N}(0,1)$, and $\alpha$ is a trade-off coefficient between the KL and MSE losses. Then, we optimize the VAE model by:
$$\theta^{\ast}_{vae} = \mathop{\arg\min}\limits_{\theta} \frac{1}{N_1}\sum_{i=1}^{N_1} \mathcal{L}(\theta_{vae}; \bar{m}_i).$$



%
%
%
%
%

\subsubsection{Input Detection.} After VAE model is fully optimized, we get optimized weights $\theta^{\ast}_{vae}$ and use the VAE test set $\{\bar{m}_i\}^{N_2}_{i = 1}$ to determine the reconstruction distance threshold $\tau|_{\mathcal{P}(dis \le \tau) = p}$ for normal samples, where the threshold $\tau$ depends on the confidence level $p$ of the benign sample reconstruction distance $dis$ and that is following $\mathcal{P}(dis \le \tau) = p$. The confidence level $p$ is set by the detector.

For an unknown input $x^{\diamond}$ to be detected, we first extract the middle feature map $\bar{m}^{\diamond}$ by $reshape(M_1( x^{\diamond}))$. Then, we can determine the inference results $Label_0$ by $M_2(\bar{m}^{\diamond})$ and $Label_1$ by $M_2(dereshape(VAE(\bar{m}^{\diamond};\theta^{\ast}_{vae})))$, where $dereshape(\cdot)$ is the reverse operation of $reshape(\cdot)$. By comparing whether $Label_0$ and $Label_1$ are equal, we can preliminarily determine whether the input carries a backdoor trigger. Then, based on whether the maximum reconstruction distance $dis$ is less than $\tau$, further judgment is made on the input $x^{\diamond}$, where $dis$ is calculated from the middle feature map $\bar{m}^{\diamond}$:
$$ res = \bar{m}^{\diamond} - VEA(\bar{m}^{\diamond}; \theta^{\ast}_{vae}), $$
$$dis=Max(| res |) - Min(| res |),$$
where $| \cdot |$ takes absolute values for all numbers in it, $Max(\cdot)$ and $Min(\cdot)$ take the maximum and minimum value respectively.
Finally, if the following conditions are met, the input samples will be recognized as backdoor samples:
$$ Label_0 \neq Label_1 \ \& \ dis > \tau.$$
Based on the Backdoor Consistency, when we detect the presence of backdoors in the input, we also believe that there are backdoors in the target DNN model and need to adopt backdoor elimination for it.
\subsection{BeniFul - Backdoor Elimination}
\begin{figure}[tbp]
    \centering
    \includegraphics[width=0.47\textwidth]{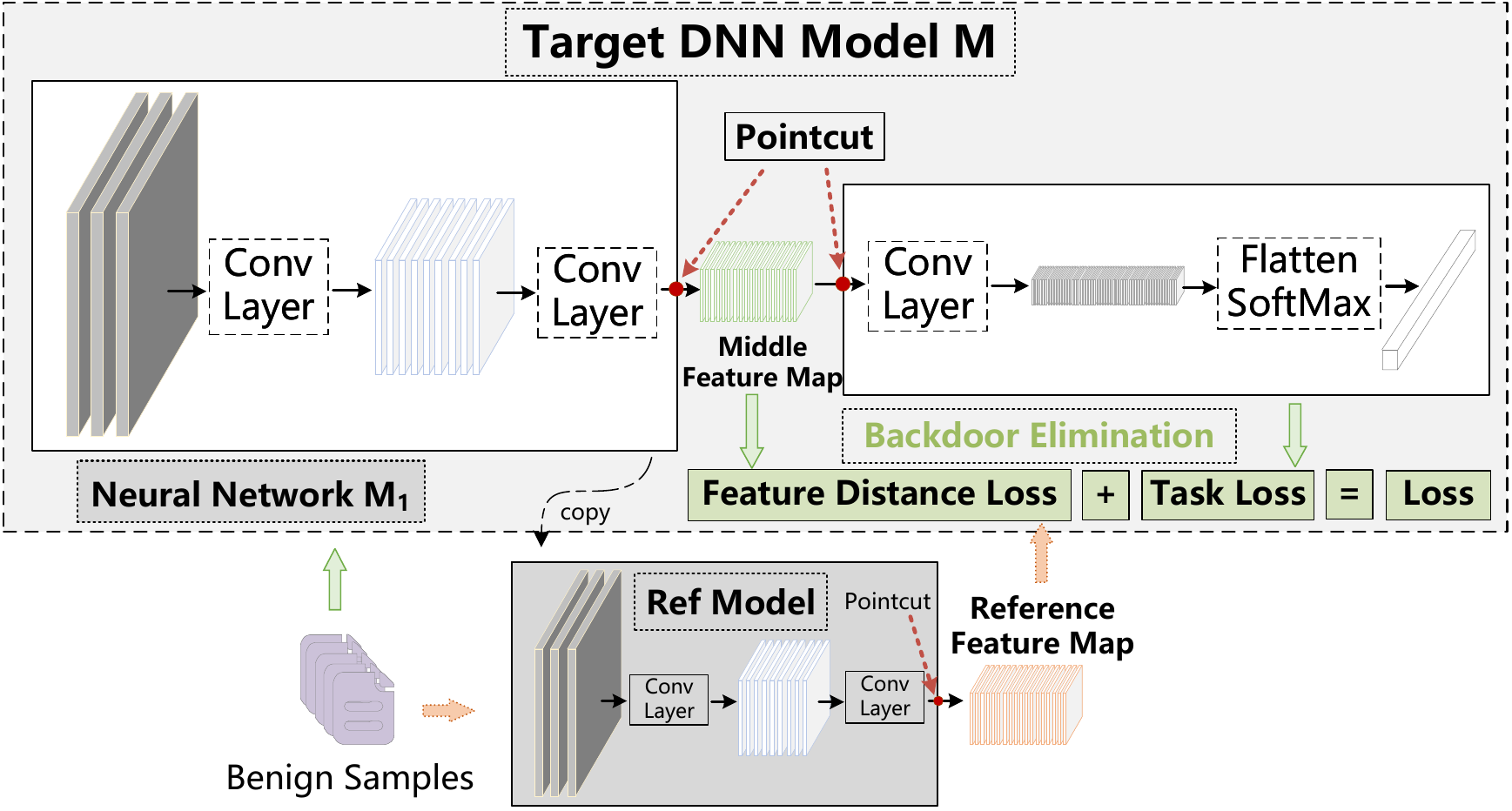}
    \caption{Backdoor Elimination.}
    \label{fig: elimination architecture}
\end{figure}

From the perspective of the middle features, we unify the features extracted from the backdoor samples into the feature space of benign samples to achieve backdoor elimination. The framework of this backdoor elimination scheme is shown in Figure~\ref{fig: elimination architecture}. We do not need to access the original backdoor samples, because we achieve backdoor elimination by a feature distance loss item that pulls the whole extracted features of the inputs away from the original features which include backdoor features. At the same time, to ensure the accuracy of the model during the elimination process, we also need to add a task loss term to the loss function.

In our backdoor elimination method, we construct a reference model $M^{ref}$ with the same structure as $M_1$ and denote the weights of the backdoor DNN model as $\theta^b$. We copy the weights of the $M_1$ model, $\theta^b_1$, to initialize the reference model weights as $\theta^{ref}$ which will never be optimized. For a sample $(x,y)$, where $y$ is the truth label of $x$, we obtain the reference feature $M^{ref}(x;\theta^{ref})$, as $m^{ref}$, which represents the original middle features for the target DNN model. Then during the elimination process, we get the middle feature $M_1(x;\theta_1^b)$, as $m^{b}$, from the poincut in the target DNN model. Then, we could obtain the Feature Distance Loss from $m^{ref}$ and $m^{b}$ of the input sample $x$:
$$ \mathcal{L}^{dis}(\theta_1^b;x, \theta^{ref}) = $$
$$-[\frac{Max(m^{b}) - m^{b}}{Max(m^{b}) - Min(m^{b})} - \frac{Max(m^{ref}) - m^{ref}}{Max(m^{ref}) - Min(m^{ref})}]^2.$$
To ensure the accuracy of the model, we need to add the Task Loss $L(\cdot)$ of the target model to the whole loss function, where Task Loss is defined as:
$$ \mathcal{L}^{task}(\theta^b;x,y) = L(M(\theta^b;x),y).$$
Then, we could get the loss function of backdoor elimination for the sample $(x,y)$:
$$ \mathcal{L}^{BE}(\theta^b;x,y,\theta^{ref}) = \mathcal{L}^{task} + \beta \cdot \mathcal{L}^{dis},$$ where $\beta$ is the trade-off coefficient between distance and task loss, and $\beta \in (0, 100]$. The backdoor elimination process for the target backdoor model in our method is shown in the Algorithm~\ref{alg: backdoor elimination}.

\section{Performance Evaluation}

\subsection{Experimental Settings}
\subsubsection{Datasets and DNN Models.}
CIFAR-10~\cite{krizhevsky2009cifar} and Tiny ImageNet~\cite{le2015tiny} are considered in our experiments. We use the ResNet-34 for these classification tasks and the pre-train model weights from PyTorch to initialize these models. During the backdoor attack simulation and backdoor elimination, we use image data augmentation which contains a random horizontal flip with a probability of $0.33$ and a random vertical flip with a probability of $0.33$.
For any image sample, we resize it to $128\times128$. All model training, testing, detecting, and eliminating processes run on one NVIDIA RTX 4090 GPU.



\begin{algorithm}[htb]
    \caption{Eliminate the backdoor in the target model.}
    \label{alg: backdoor elimination}
    \textbf{Input}: Clean sample set $\{(x_i, y_i)\}^{N}_{i=1}$, target backdoor model $M(x;\theta^b)$, reference model $M^{ref}(x;\theta^{ref})$, learning rate $lr$, batch size $n$.\\
    \textbf{Output}: Non-backdoor weights $\theta$.

    \begin{algorithmic}[1] 
    \STATE Initialize the weights $\theta^{ref}$ by $\theta^b_1$

    \FOR {$each \ epoch$}
        \FOR {$i = 1,\ 2, \ ... \ \lfloor \frac{N}{n} \rfloor $}
            \STATE $loss = \frac{1}{n}\sum_{j=(i-1)\cdot n + 1}^{i \cdot n} \mathcal{L}^{BE}(\theta^{b}; x_j, y_j, \theta^{ref}) $
            \STATE $\theta^{b} \leftarrow \theta^{b} - lr \cdot \bigtriangledown_{\theta} loss$
        \ENDFOR
    \ENDFOR
    \STATE $\theta \leftarrow \theta^{b}$
    \STATE \textbf{return} $\theta$
    \end{algorithmic}
\end{algorithm}

\subsubsection{Evaluation Metrics.} We use three metrics to evaluate the performance of our methods: Clean Accuracy (ACC), Area Under the Receiver Operating Characteristic (AUROC), and Attack Success Ratio (ASR).
For backdoor detection, we use AUROC to evaluate detection schemes. If the AUROC score of the scheme is higher, it indicates that the scheme is more effective in distinguishing normal samples and backdoor samples. For backdoor elimination, we use ACC and ASR to evaluate elimination schemes. During backdoor elimination, the scheme has better effectiveness if it can reduce a higher ASR with fewer ACC.

\subsubsection{Attack Baseline.} 
\begin{figure}[h]
    \centering
    \includegraphics[width=0.45\textwidth]{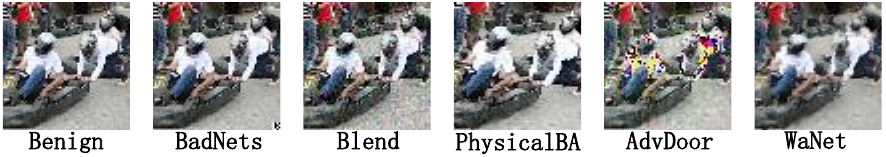}
    \caption{Benign and Backdoor Images of Tiny ImageNet.}
    \label{fig: backdoor trigger}
\end{figure}

Five state-of-the-art(SOTA) backdoor attack methods are considered in our experiments, which are BadNets\cite{Gu2019BadNets}, Blend\cite{Chen2017Targeted}, PhysicalBA\cite{Wenger2021PhyasicalBA}, AdvDoor\cite{Zhang2021AdvDoor}, and WaNet\cite{Nguyen2021WaNet}. For BadNets, we modify the value of $9$ pixels in the bottom right corner of the images in the training set to construct backdoor images. For Blend, we blend the image with Gaussian noise. For PhysicalBA, we randomly change the brightness and contrast of the image, and make random affine transformation. For AdvDoor, we train a clean model, generate adversarial perturbations as the backdoor trigger, and add this trigger to the images. For WaNet, we warp the image following the method in the original paper. The benign and each backdoor images are shown in Figure~\ref{fig: backdoor trigger} and the performance of each backdoor attack method, as the attack baseline of our experiments, is shown in the ``Non" column of Table~\ref{tab: elimination comparison}. The default poisoning ratio in our experiments is set to $10\%$.

\subsubsection{Comparison Mechanisms.} For backdoor detection, we compare our detection method with $4$ SOTA methods: STRIP~\cite{Gao2019STRIP}, FreqDetector~\cite{Zeng2021FreqDetector}, SCALE-UP~\cite{Guo2023SCALE}, and TeCo~\cite{Liu2023TeCo}. For backdoor elimination, we compare our elimination method with $4$ SOTA methods: FP~\cite{Liu2018FP}, NAD~\cite{Li2021NAD}, ABL~\cite{Li2021ABL}, and CBD~\cite{Zhang2023CBD}.

\subsection{Experiments}
\subsubsection{Effectiveness of Our Method.}
To verify the effectiveness of our mechanism, we evaluate our own backdoor detection and elimination method on Tiny ImageNet datasets under five attack methods as shown in Table~\ref{tab: attack baseline and effectiveness}. The architecture of the VAE model in our detection method is shown in Appendix~B, and we choose the pointcut between the third and the fourth residual block of ResNet-34. In our detection experiments, we set default $\alpha$ as $0.5$ in the loss function and use the middle features from $15\%$ clean training set to train the VAE model.
In our elimination experiments, we set default $\beta$ as $35$ and use $15\%$ clean training set to eliminate the backdoor. The column ``Detection'' in Table~\ref{tab: attack baseline and effectiveness} corresponds to our detection scheme. The AUROC index of our detection method could reach at least $92\%$. The column ``Elimination'' indicates the result after backdoor elimination using our elimination method. For different attacks, our approach can eliminate backdoors in these models, in which the ASR has decreased by more than $95\%$, and maintain the accuracy loss within $5\%$ in most cases.

\begin{table}[h]
\setlength\tabcolsep{3.8pt}
\centering
\begin{tabular}{c|cc|c|cc}
\hline
\multirow{2}{*}{Attack Method} & \multicolumn{2}{c|}{Non} & \textbf{Detection} & \multicolumn{2}{c}{\textbf{Elimination}}\\
 &   ACC & ASR &  AUROC& ACC & ASR \\
\hline
BadNets &0.692 &0.989 &0.958 &0.675 &0.027\\

Blend &0.696 &0.998 &0.979 &0.641 &0.014\\

PhysicalBA &0.689 &0.998 &0.921 &0.664 &0.026\\

AdvDoor &0.678 &0.999 &0.978 &0.643 &0.034\\

WaNet &0.671&0.973&0.928 &0.658 &0.021\\
\hline
\end{tabular}
\caption{Effectiveness of Our Methods.}
\label{tab: attack baseline and effectiveness}
\end{table}

\begin{figure}[h]
\centering
    \begin{subfigure}[t]{0.32\columnwidth}
  		\centering
  		\includegraphics[width=\linewidth]{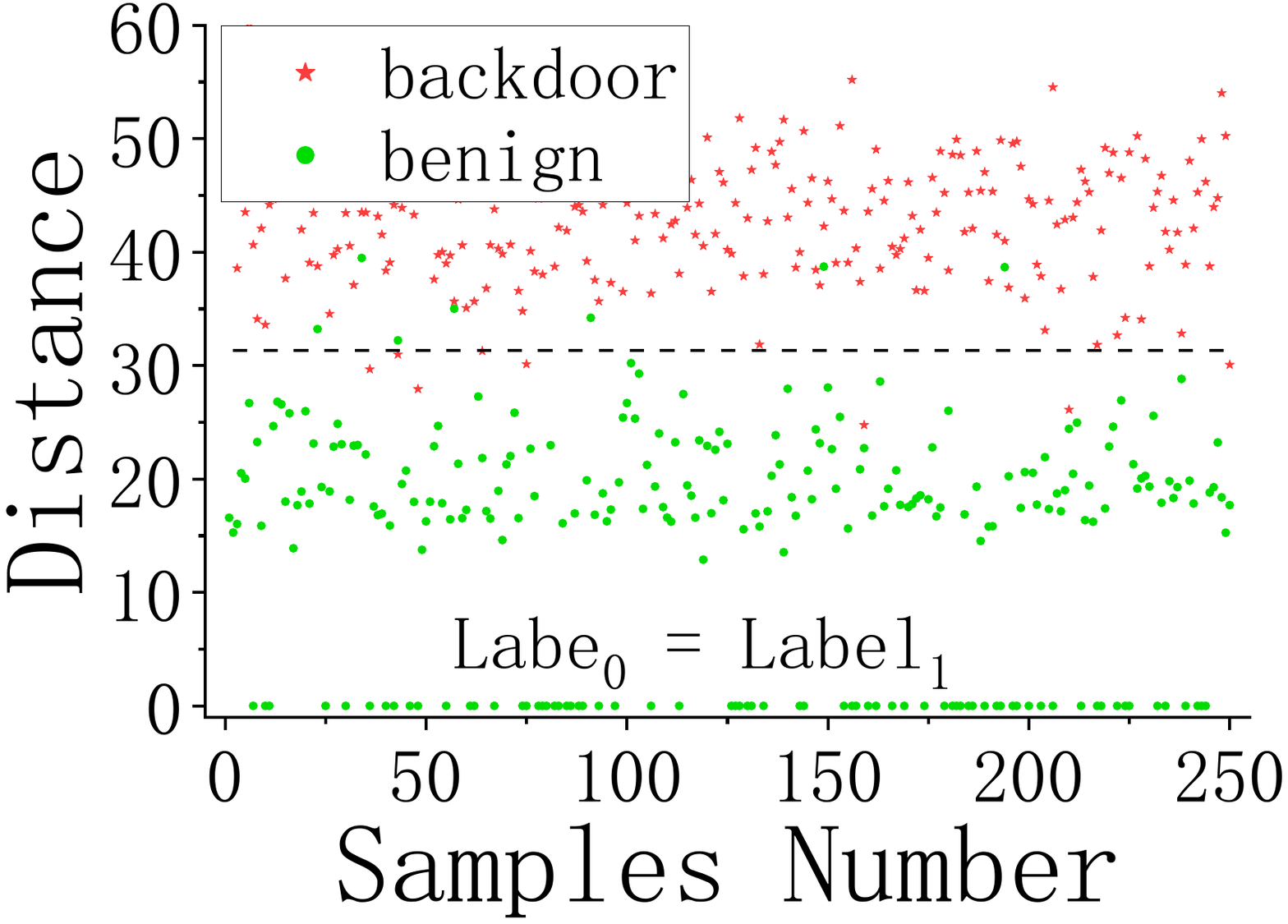}
  		\caption{Blend}
        \label{fig: scatter blend}
    \end{subfigure}
    \begin{subfigure}[t]{0.32\columnwidth}
  		\centering
  		\includegraphics[width=\linewidth]{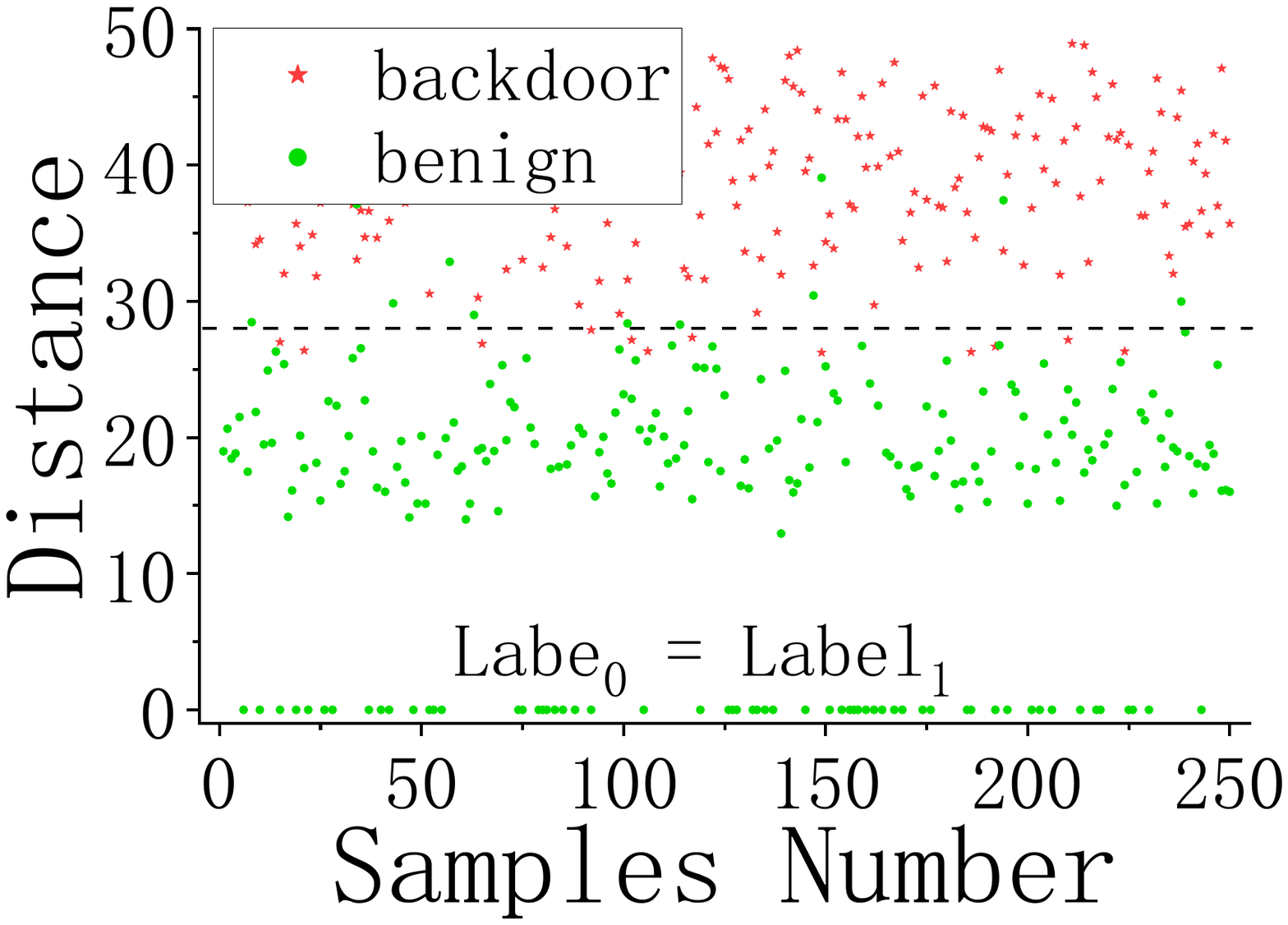}
  		\caption{AdvDoor}
    \end{subfigure}
    \begin{subfigure}[t]{0.32\columnwidth}
  		\centering
  		\includegraphics[width=\linewidth]{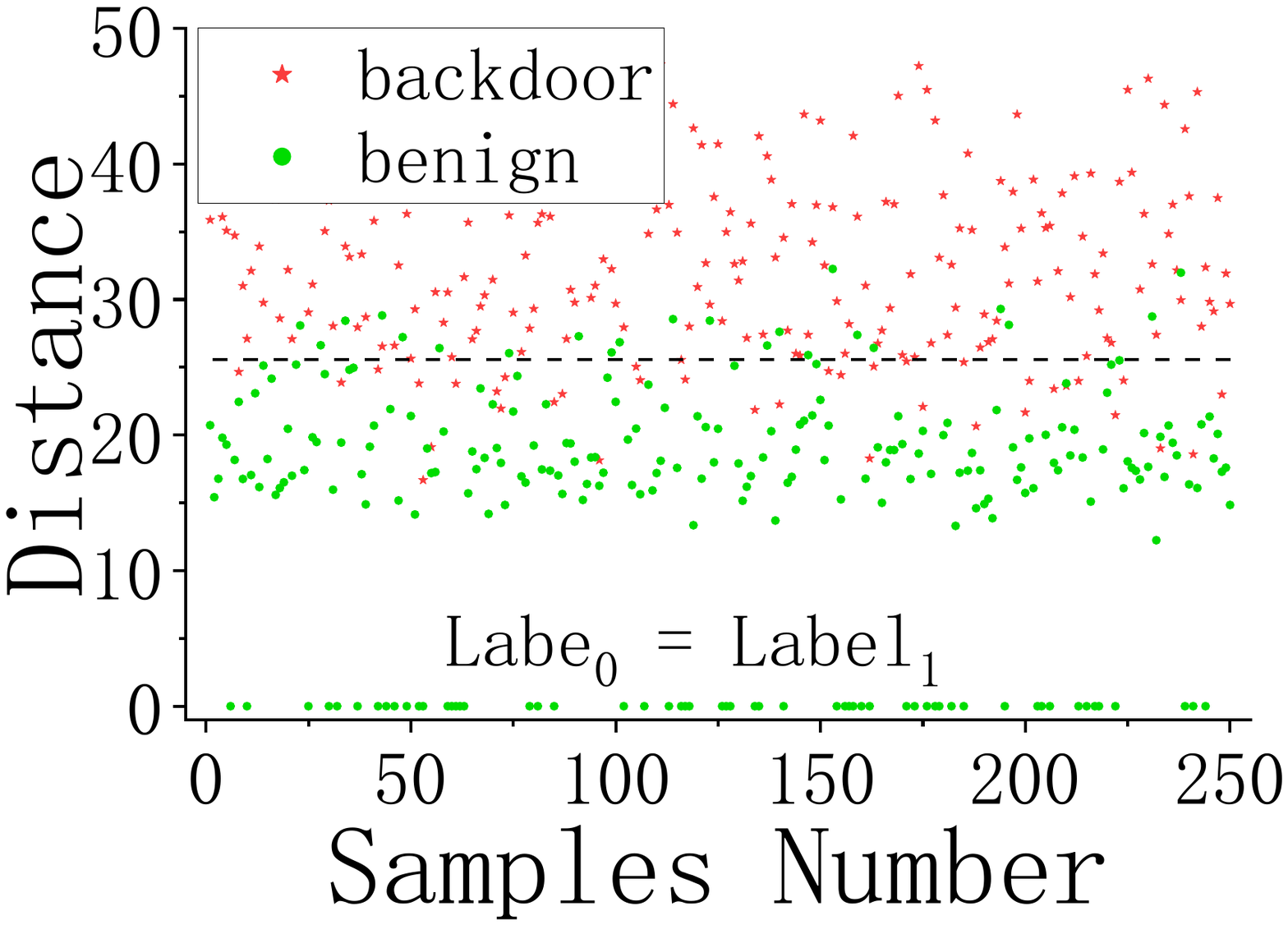}
  		\caption{WaNet}
    \end{subfigure}
\caption{Backdoor Detection on Tiny ImageNet.}
\label{fig: scatter}
\end{figure}

In order to have a more intuitive understanding of the experimental results of our detection method, we randomly select $250$ images from the test set of Tiny ImageNet and draw a binary scatter plot to reconstruct the distance based on whether to add the Blend, AdvDoor, or WaNet backdoor trigger to these images, as shown in Figure~\ref{fig: scatter}. For an example of the Blend Attack, as shown in Figure~\ref{fig: scatter}(a), we set distance as $0$ when the $Label_1$ is equal to $Label_0$ and we could choose the distance of $31.8$ as $\tau$ to achieve a great backdoor input detection with $97.9\%$ AUROC score.

\begin{figure}[h]
\centering
    \begin{subfigure}[t]{0.32\columnwidth}
  		\centering
  		\includegraphics[width=\linewidth]{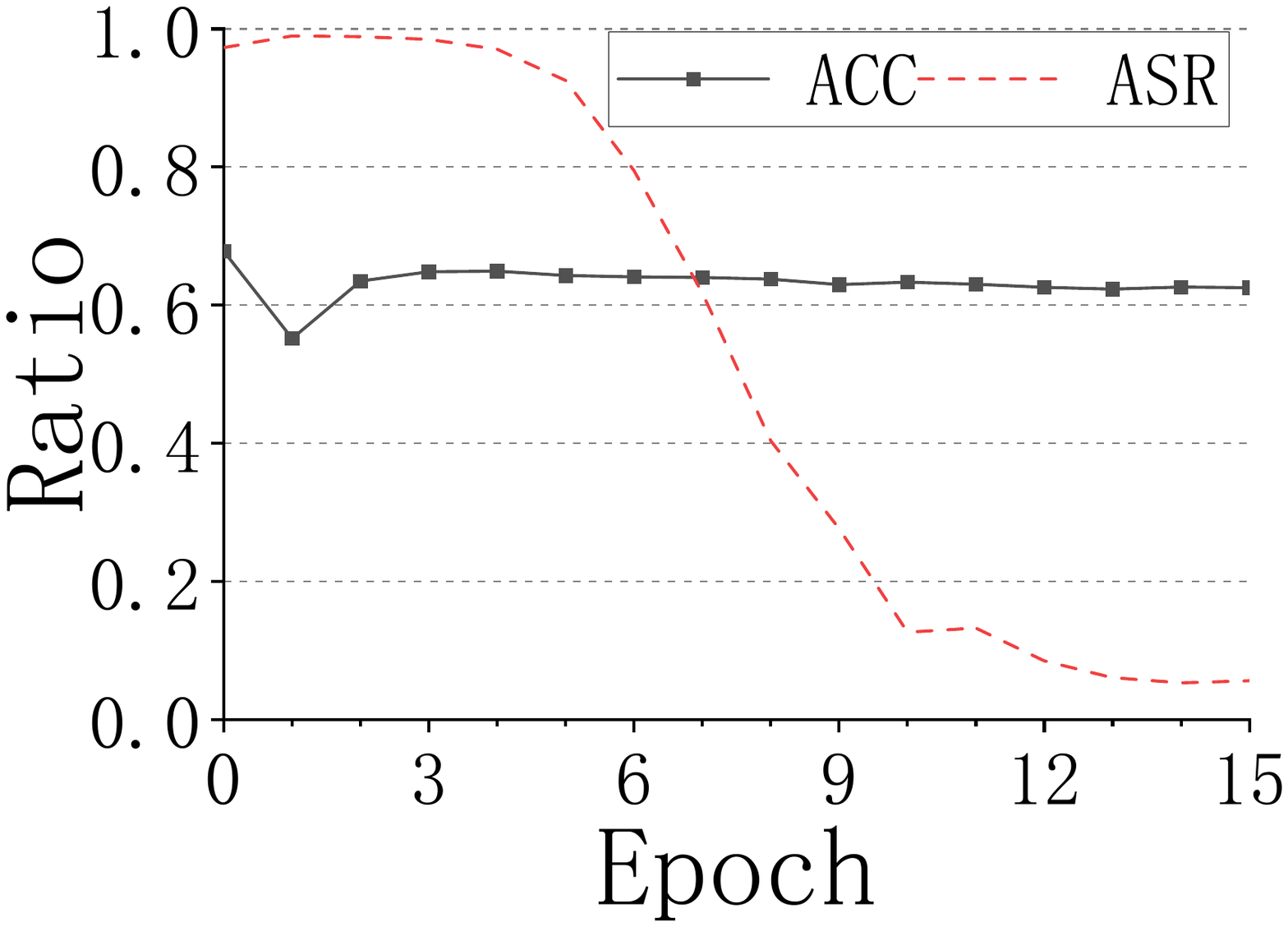}
  		\caption{Blend}
    \end{subfigure}
    \begin{subfigure}[t]{0.32\columnwidth}
  		\centering
  		\includegraphics[width=\linewidth]{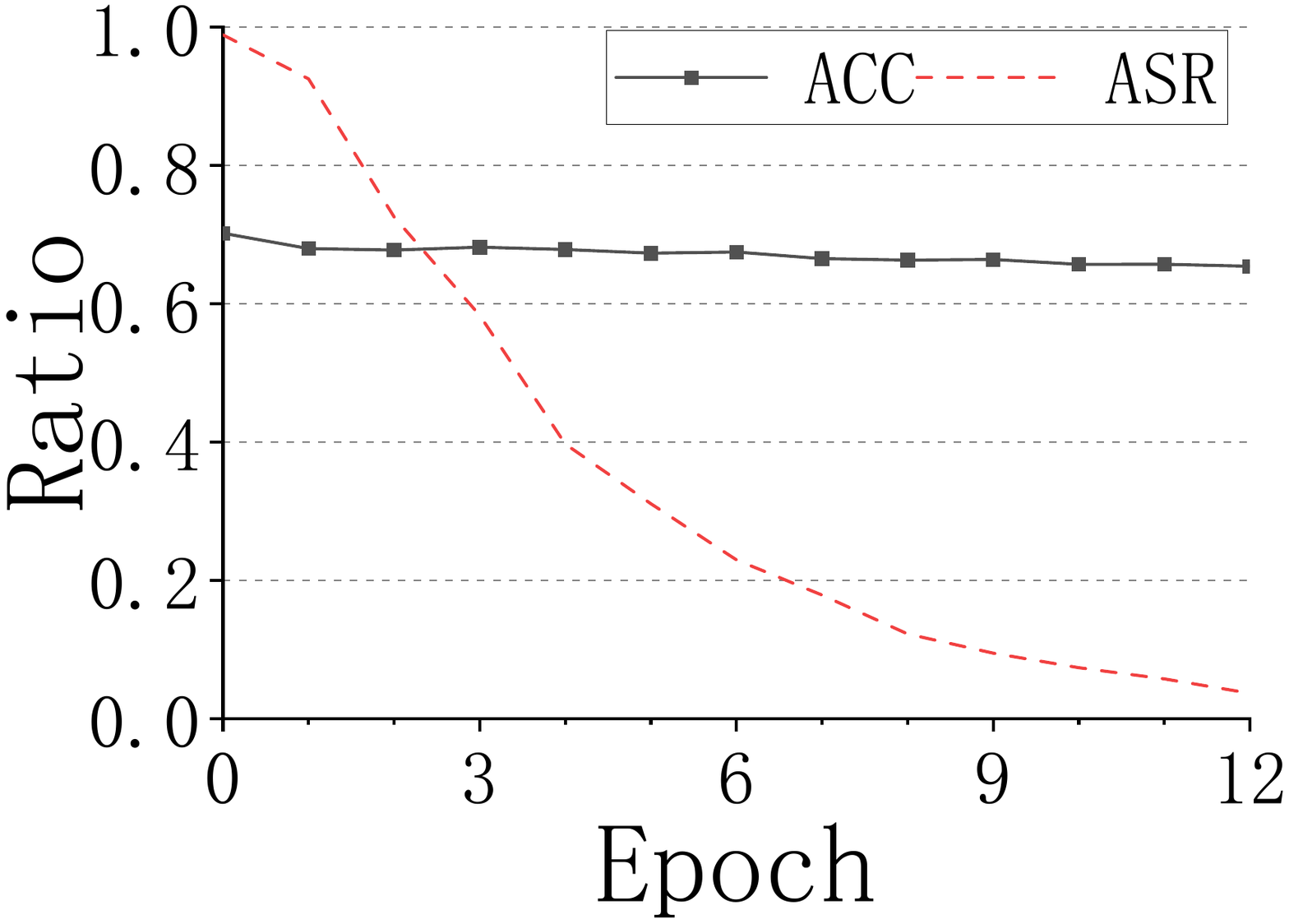}
  		\caption{AdvDoor}
    \end{subfigure}
    \begin{subfigure}[t]{0.32\columnwidth}
  		\centering
  		\includegraphics[width=\linewidth]{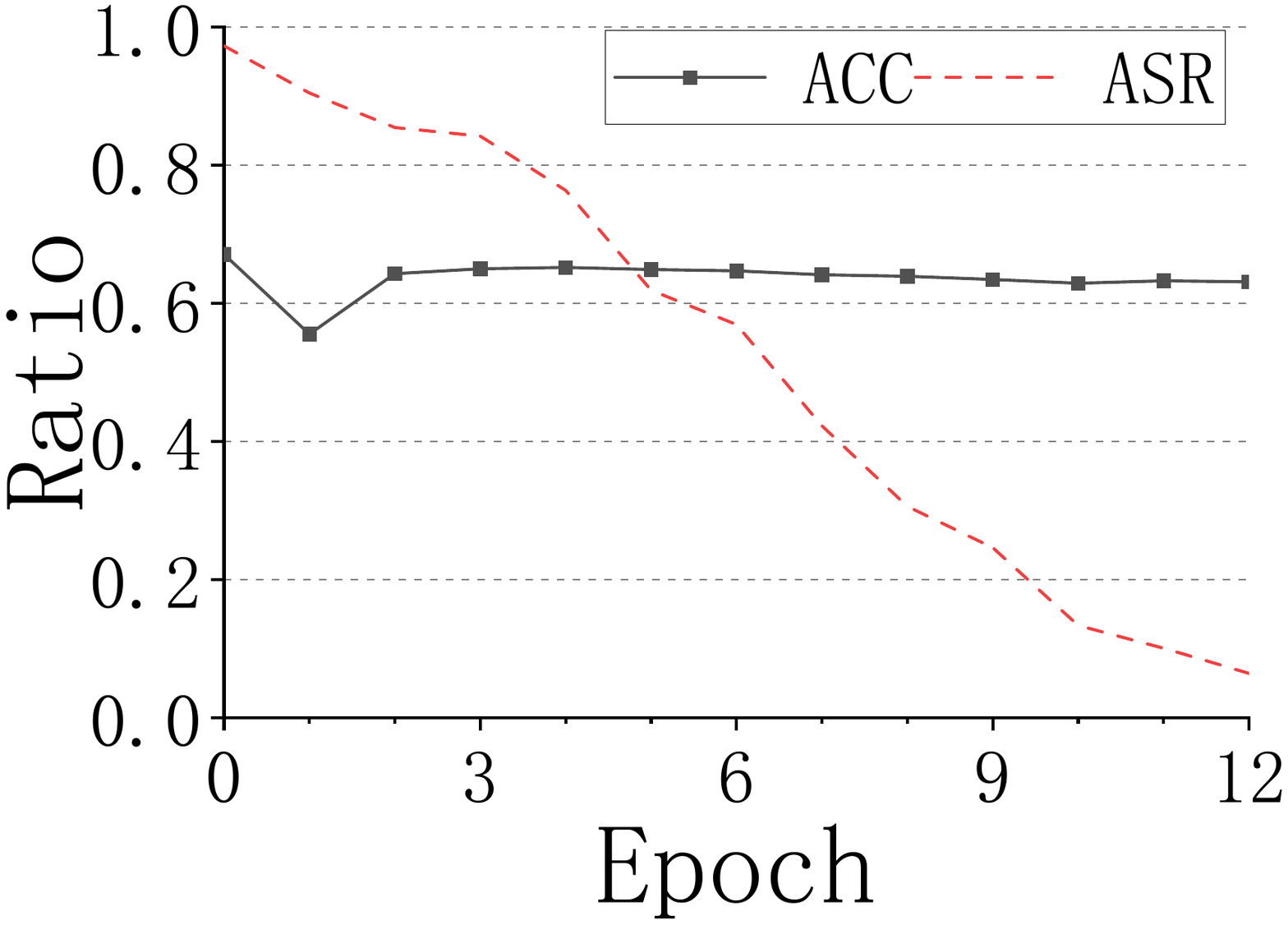}
  		\caption{WaNet}
    \end{subfigure}
\caption{Backdoor Elimination on Tiny ImageNet.}
\label{fig: elimination process}
\end{figure}

To better observe the backdoor elimination process, we draw the curves of ACC and ASR during the process of eliminating the backdoor. As shown in Figure~\ref{fig: elimination process} on the Tiny ImageNet dataset, for Blend, AdvDoor, and WaNet attack methods, our method could eliminate the backdoor within $15$ epoch and retain the ACC upon $64\%$.

\begin{table*}[t]
\setlength\tabcolsep{11pt}
\centering
\begin{tabular}{c|c|cccccc}
\hline
\multirow{2}{*}{Dataset} & Attack$\rightarrow$ & \multirow{2}{*}{BadNets} & \multirow{2}{*}{Blend} & \multirow{2}{*}{PhysicalBA} & \multirow{2}{*}{AdvDoor} & \multirow{2}{*}{WaNet} & \multirow{2}{*}{$Average$}\\
& Detection$\downarrow$ &  &  &  &  &  & \\
\hline
\multirow{5}{*}{CIFAR-10}
 & STRIP &0.978 & 0.933 & 0.938 & 0.752 & 0.567 &0.834\\
 & FreqDetector & 0.897 & 0.979 & 0.838 &0.883 & 0.588&0.837\\
 & SCALE-UP & 0.965 & 0.958 & 0.935 & 0.847 & 0.901&0.921\\
 & TeCo & 0.882 & 0.928 & 0.894 & 0.887 & 0.896&0.897\\
 & \textbf{Ours} & 0.986 & 0.988 & 0.954 & 0.921 & 0.964&0.962\\
\hline
\multirow{5}{*}{Tiny ImageNet}
 & STRIP & 0.928 & 0.952 & 0.947 &0.718 &0.599&0.829\\
 & FreqDetector & 0.859 & 0.938 &0.882 &0.819 &0.628&0.825\\
 & SCALE-UP & 0.935 & 0.949 & 0.909 & 0.742 &0.925&0.892\\
 & TeCo & 0.943 & 0.937 & 0.861 & 0.854 &0.917 &0.902\\
 & \textbf{Ours} & 0.958 & 0.979 & 0.921 & 0.978 & 0.928 &0.953\\
\hline
\end{tabular}
\caption{Comparison of Backdoor Input Detection Methods.}
\label{tab: detection comparison}
\end{table*}

\begin{table*}[t]
\setlength\tabcolsep{4pt}
\centering
\begin{tabular}{c|c|cccccccccccc}
\hline
 \multirow{2}{*}{Dataset} & Elimination $\rightarrow$ & \multicolumn{2}{c}{Non} & \multicolumn{2}{c}{FP} & \multicolumn{2}{c}{NAD} & \multicolumn{2}{c}{ABL} & \multicolumn{2}{c}{CBD} & \multicolumn{2}{c}{\textbf{Ours}} \\
 & Attack$\downarrow$ & ACC & ASR & ACC & ASR & ACC & ASR & ACC & ASR & ACC & ASR & ACC & ASR\\
\hline
\multirow{6}{*}{CIFAR-10}
 & BadNets      &0.941&0.998 &0.912&0.756 &0.854&0.036 &0.866&0.096 &0.878&0.048 &0.925 &0.028\\
 & Blend        &0.948&0.999 &0.903&0.872 &0.849&0.063 &0.843&0.163 &0.856&0.035 &0.913 &0.043\\
 & PhysicalBA   &0.914&0.979 &0.883&0.856 &0.835&0.069 &0.798&0.133 &0.882&0.063 &0.915 &0.053\\
 & AdvDoor      &0.945&0.983 &0.898&0.912 &0.847&0.083 &0.826&0.096 &0.869&0.088 &0.906 &0.071\\
 & WaNet        &0.934&0.999 &0.853&0.814 &0.878&0.104 &0.861&0.089 &0.868&0.064 &0.911 &0.040\\
 & $Average$    &0.936&0.992	&0.890	&0.842	&0.853	&0.071	&0.839	&0.115	&0.871	&0.060	&0.914	&0.047	\\
\hline
\multirow{6}{*}{Tiny ImageNet}
 & BadNets      &0.692&0.989 &0.647&0.895 &0.667&0.033 &0.641&0.083 &0.652&0.071 &0.678 &0.027\\
 & Blend        &0.696&0.998 &0.658&0.903 &0.653&0.078 &0.634&0.257 &0.643&0.067 &0.641 &0.014\\
 & PhysicalBA   &0.689&0.998 &0.633&0.828 &0.643&0.083 &0.655&0.171 &0.659&0.098 &0.664 &0.026\\
 & AdvDoor      &0.678&0.999 &0.646&0.859 &0.656&0.068 &0.649&0.099 &0.648&0.082 &0.643 &0.034\\
 & WaNet        &0.671&0.973 &0.635&0.813 &0.644&0.128 &0.638&0.076 &0.647&0.049 &0.658 &0.021\\
 & $Average$    &0.685	&0.991	&0.644	&0.860	&0.653	&0.078	&0.643	&0.137	&0.650	&0.073	&0.657	&0.024\\
\hline
\end{tabular}
\caption{Comparison of Backdoor Elimination Methods.}
\label{tab: elimination comparison}
\end{table*}

\subsubsection{Comparison with Existing Methods.} Table~\ref{tab: detection comparison} shows the comparing result between our detection method and the four existing SOTA backdoor detection methods for five backdoor attacks and on CIFAR-10 and Tiny ImageNet datasets. For STRIP, FreqDetector, SCALE-UP, and TeCo, we comply with the settings as their original papers. STRIP and FreqDetector make it hard to detect the WaNet backdoor inputs due to the disadvantages of these schemes. For example, STRIP which detected backdoor inputs by adding strong intentional perturbation has good performance for BadNets, Blend, or other patch-based backdoor attacks, because their triggers are not related to the image features. Therefore, it has less advantages of detecting transform-based backdoor attacks, like WaNet. SCALE-UP detected backdoor inputs by the scaled prediction consistency (SPC) and Teco was by the corruption robustness consistency (CRC). These two backdoor detection methods are effective in these five backdoor attacks. However, when these two schemes perform backdoor detection on input, they require multiple instances of varying degrees of sample perturbation or pixel scale-up, and the model needs multiple inferences on this input as shown in Table~\ref{tab: inference times}, which reduces detection efficiency even model runs efficiency. We refer to this kind of approach as the Inference-Consistency-based method, which detects by modifying input samples multiple times and determining whether an input is a backdoor based on the consistency of the model's multiple inference results.

\begin{table}[!htbp]
\centering
\begin{tabular}{ccc}
\hline
Method & Inference Times & Basic Principle \\
\hline
STRIP & 100 &  Inference Consistency\\
FreDetector & 1 & Frequency \\
SCALE-UP & 10 & Inference Consistency \\
TeCo & $80$ & Inference Consistency\\
Our & 1 & Middle Feature \\
\hline
\end{tabular}
\caption{Inference Times and Its Detection Principle.}
\label{tab: inference times}
\end{table}

Compared with $4$ existing SOTA elimination methods, including FP, NAD, ABL, and CBD, our backdoor elimination method achieves more ASR reduction and less ACC loss as shown in Table~\ref{tab: elimination comparison}. FP eliminates backdoors by pruning and fine-tuning the backdoor model. Thus, it is difficult to completely eliminate the backdoors of the model. NAD implements backdoor elimination through knowledge distillation, where the teacher model is clean and takes the backdoor model as the student model. The model's accuracy after elimination is largely dependent on the performance of the teacher model. ABL and CBD achieved backdoor defense by training a clean model on a backdoor dataset, but the accuracy of the clean model might be affected. In contrast, our method eliminates the backdoor in the already trained backdoor model without introducing additional DNN models.

\subsection{Backdoor Consistency}\label{sec: backdoor consistency}
To understand the backdoor attack from the perspective of the middle feature, which comes from the middle layer output of the DNN model, we use the CIFAR100~\cite{krizhevsky2009cifar} dataset and the ResNet-34 model to conduct the following inspiring experiments. The data augmentation follows our above experimental settings. To search for the model inference differences from the middle features between benign and backdoor samples, we drop out the feature outputs from the fourth residual block of ResNet-34 and obtain the difference in the inference results between benign and backdoor samples with varying dropout rates (DR). As shown in Appendix~A, when we set $95\%$ dropout ratio for the middle features, the ACC of the model significantly decreases to less than $35\%$, while the ASR of BadNets, WaNet, and AdvDoor remains high at more than $90\%$.


\begin{figure}[h]
\centering
\includegraphics[width=\columnwidth]{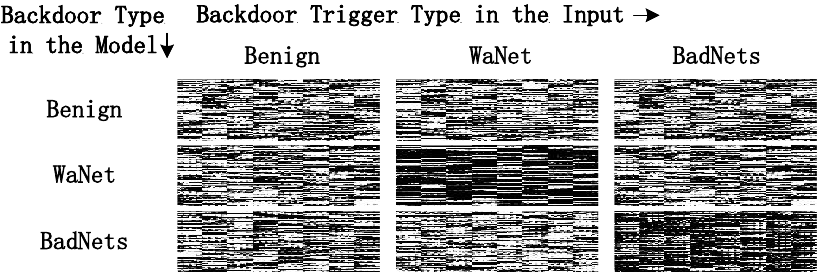}
\caption{Comparison of Different Middle Features.}
\label{fig: middle feature}
\end{figure}


This phenomenon encourages us to make further exploration of the relationship between middle features and backdoors. For each input where the middle features are $256$ channels and $4 \times 4$ feature map of each channel, we reshape the middle features into a grayscale image shaped as $1 \times 32 \times 64$. We find a certain difference between the features extracted by the model from normal samples and backdoor samples. For example, as shown in Figure~\ref{fig: middle feature}, in the backdoor model that suffers from the WaNet attack, the middle features extracted from the WaNet backdoor input differ from the middle features of both benign and BadNet backdoor inputs, indicating that the backdoor input has a strong consistent association with the backdoor model and this inspires us to achieve backdoor defense from the middle features.

\subsection{Ablation Studies}
\subsubsection{Impact of the Hyperparameter $\alpha$.}
\begin{figure}[h]
\centering
    \begin{subfigure}[t]{0.32\columnwidth}
  		\centering
  		\includegraphics[width=\linewidth]{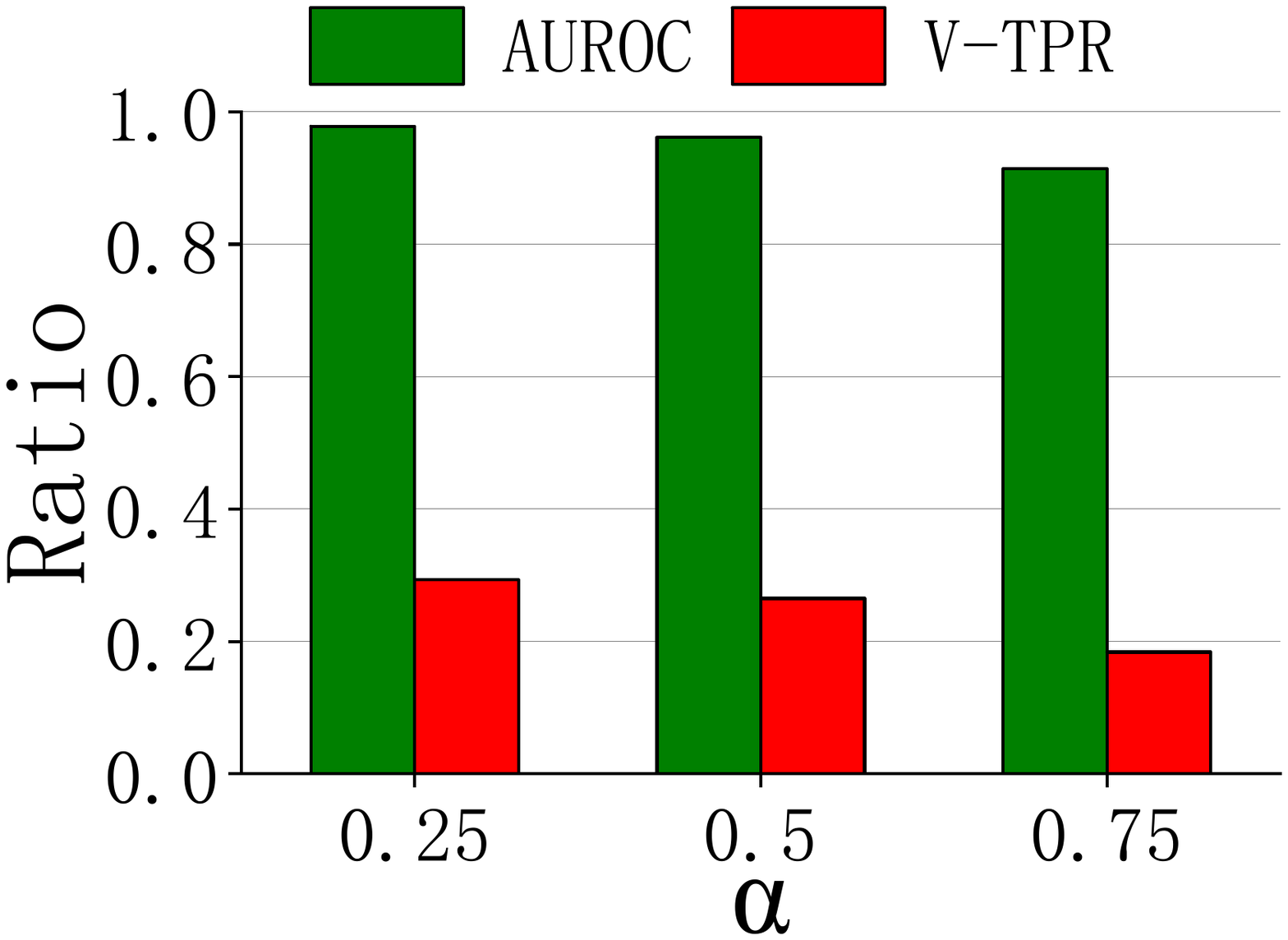}
  		\caption{Blend}
    \end{subfigure}
    \begin{subfigure}[t]{0.32\columnwidth}
  		\centering
  		\includegraphics[width=\linewidth]{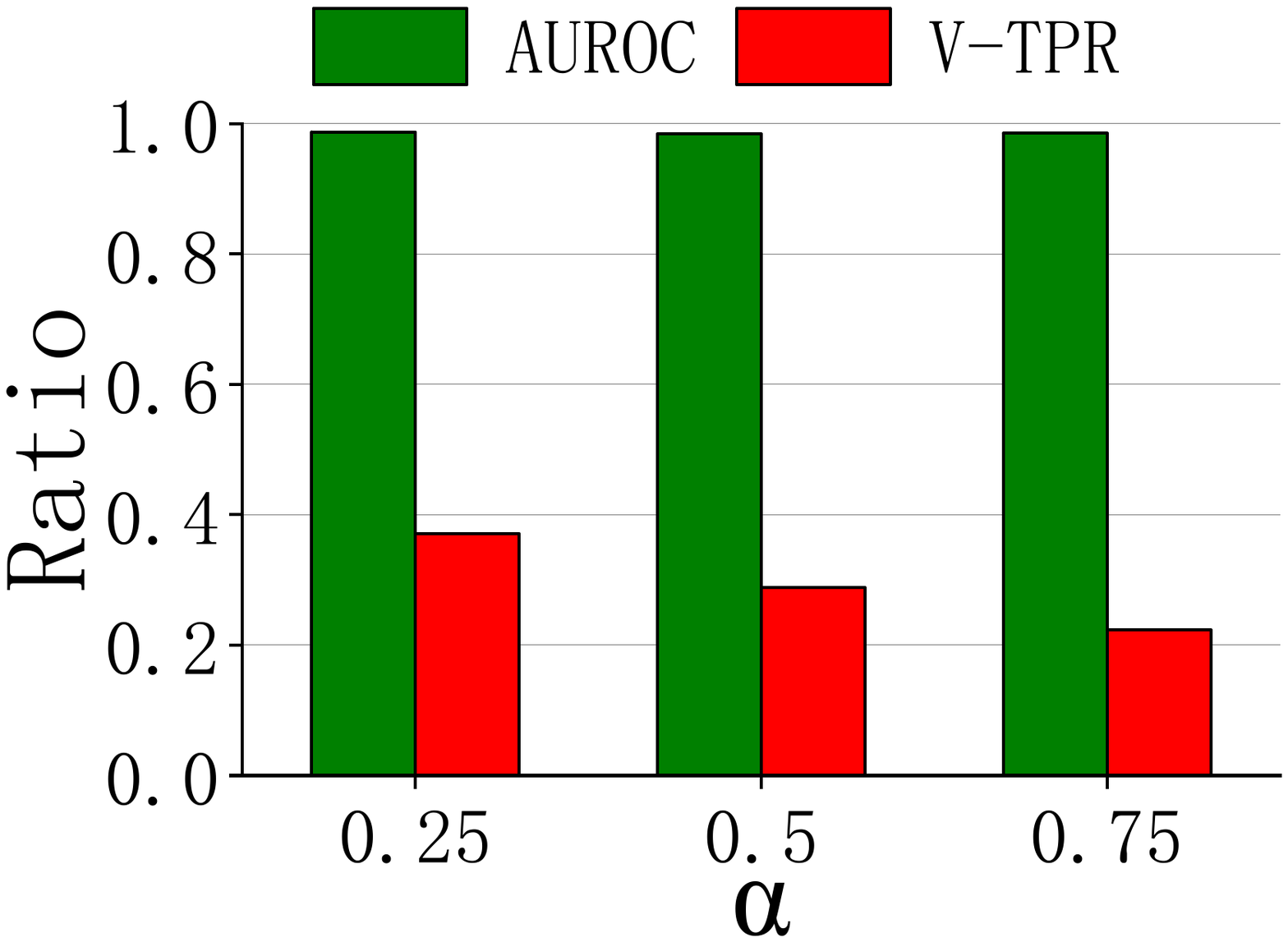}
  		\caption{AdvDoor}
    \end{subfigure}
    \begin{subfigure}[t]{0.32\columnwidth}
  		\centering
  		\includegraphics[width=\linewidth]{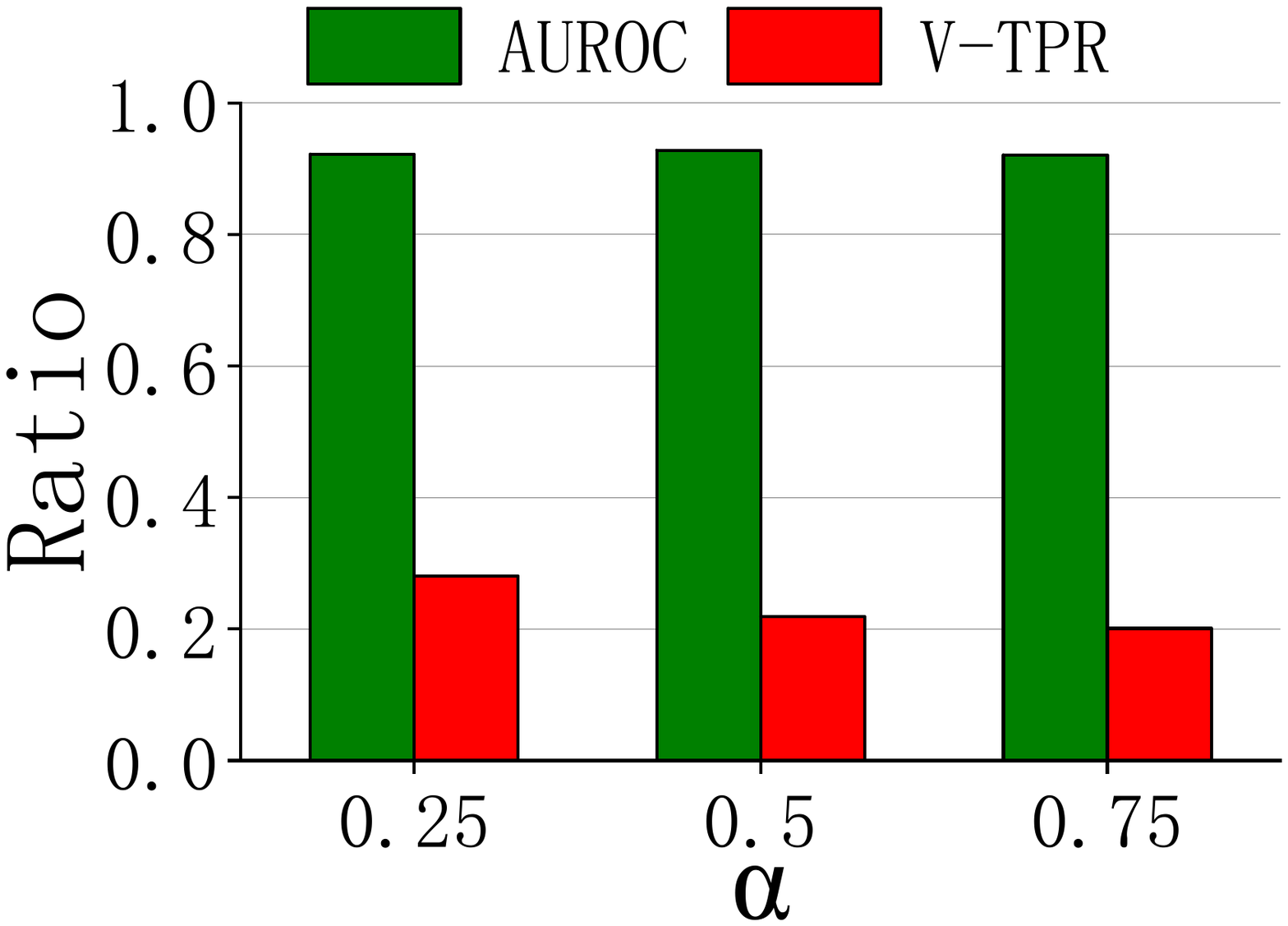}
  		\caption{WaNet}
    \end{subfigure}
\caption{Detection Performance with Different $\alpha$.}
\label{fig: alpha}
\end{figure}

The $\alpha$ is the trade-off coefficient between KL loss and reconstruction loss. To evaluate the impact of this hyperparameter on the detection result, we design two items to reveal this impact: the detection result which is reflected by the detection AUROC score and the positive feature reconstruction ability of VAE which is reflected by the ratio of benign test samples with $Label_0 = Label_1$ to all benign test samples during detection, denoted as V-TPR. On the Tiny ImageNet dataset, as shown in Figure~\ref{fig: alpha}, $\alpha$ greatly impacts the positive feature reconstruction ability but has only a minor impact on the detection result for different backdoor attacks.

\subsubsection{Impact of the Hyperparameter $\beta$.} 
\begin{figure}[h]
\centering
    \begin{subfigure}[t]{0.32\columnwidth}
  		\centering
  		\includegraphics[width=\linewidth]{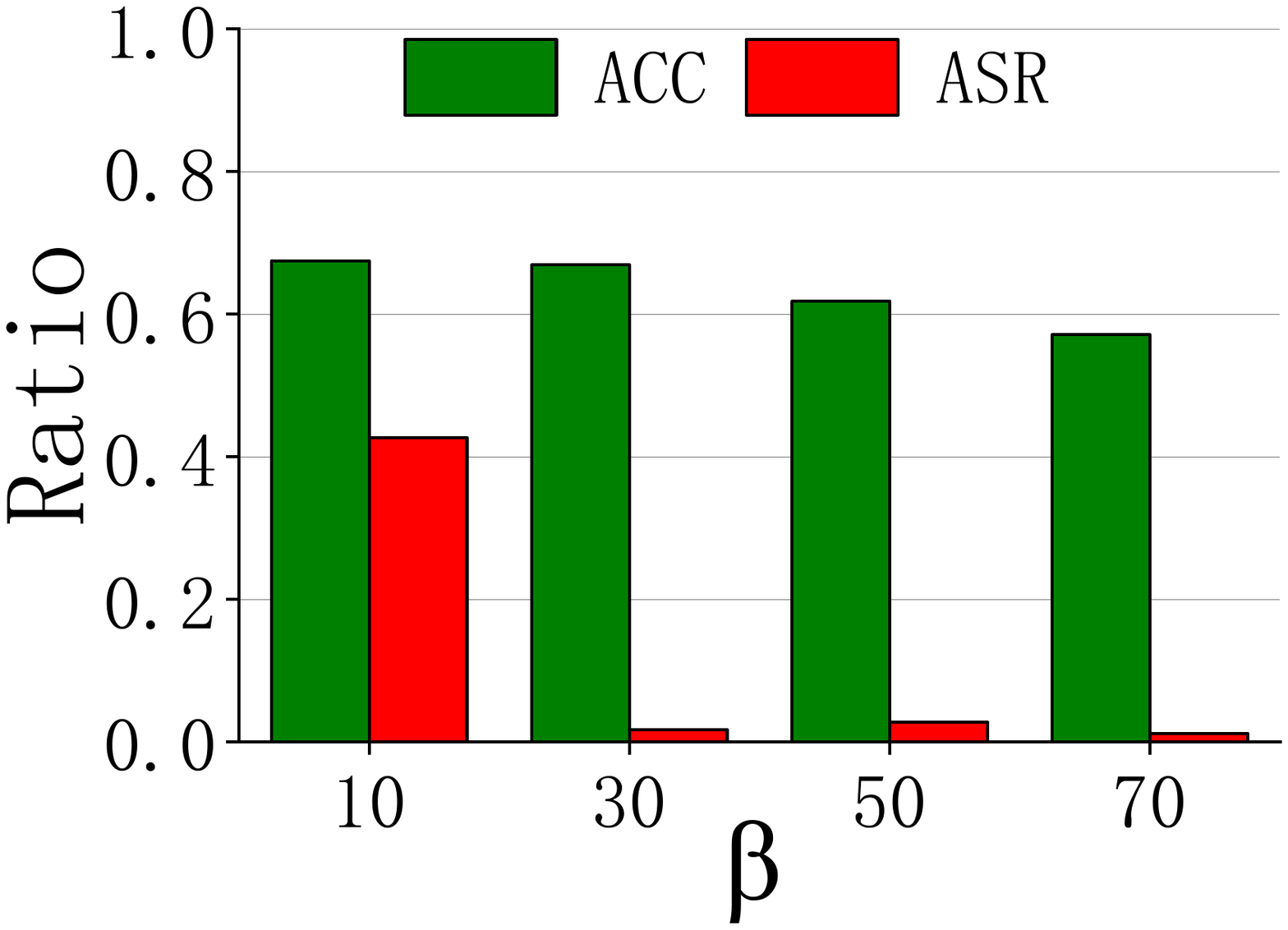}
  		\caption{Blend}
    \end{subfigure}
    \begin{subfigure}[t]{0.32\columnwidth}
  		\centering
  		\includegraphics[width=\linewidth]{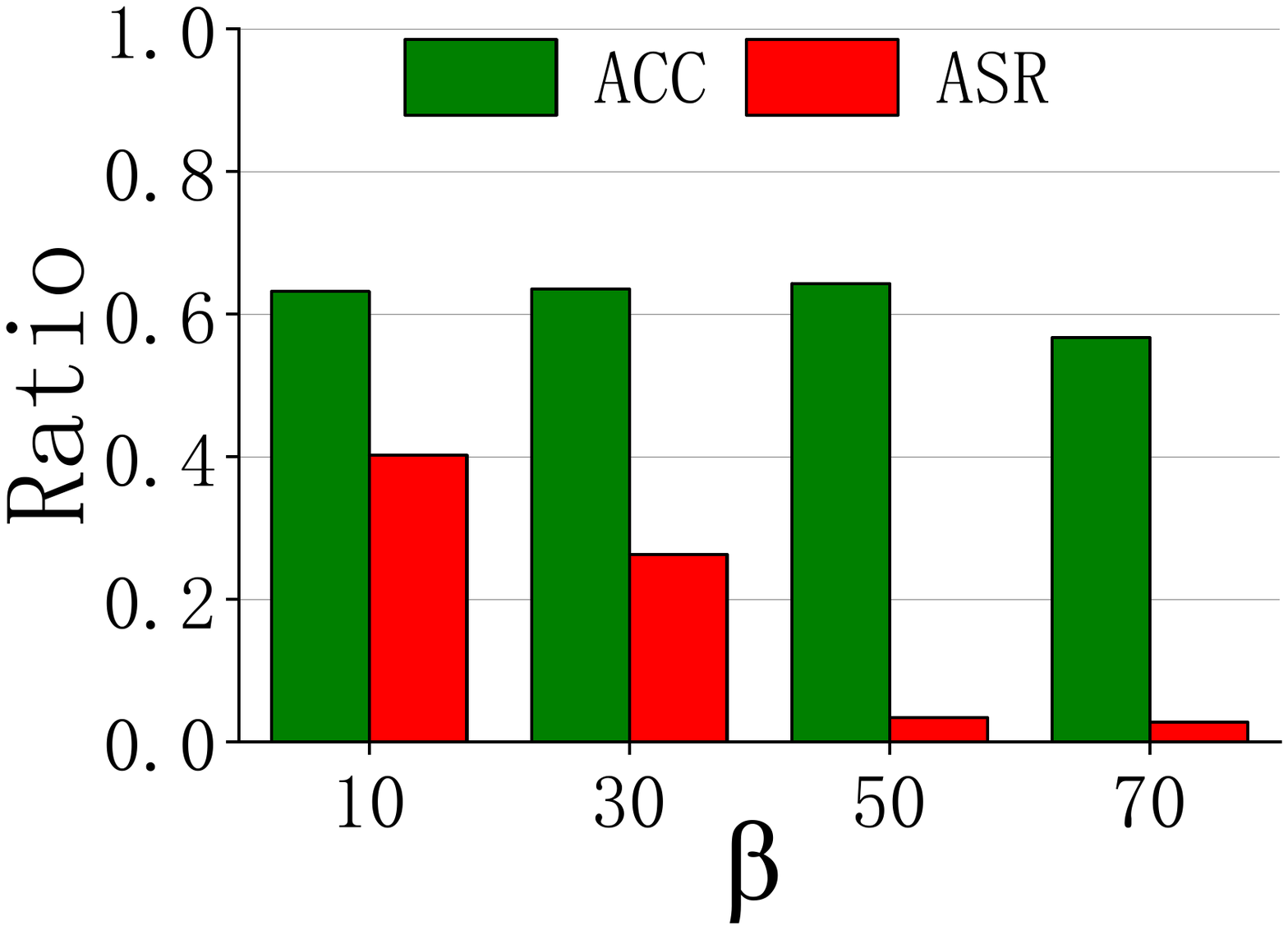}
  		\caption{AdvDoor}
    \end{subfigure}
    \begin{subfigure}[t]{0.32\columnwidth}
  		\centering
  		\includegraphics[width=\linewidth]{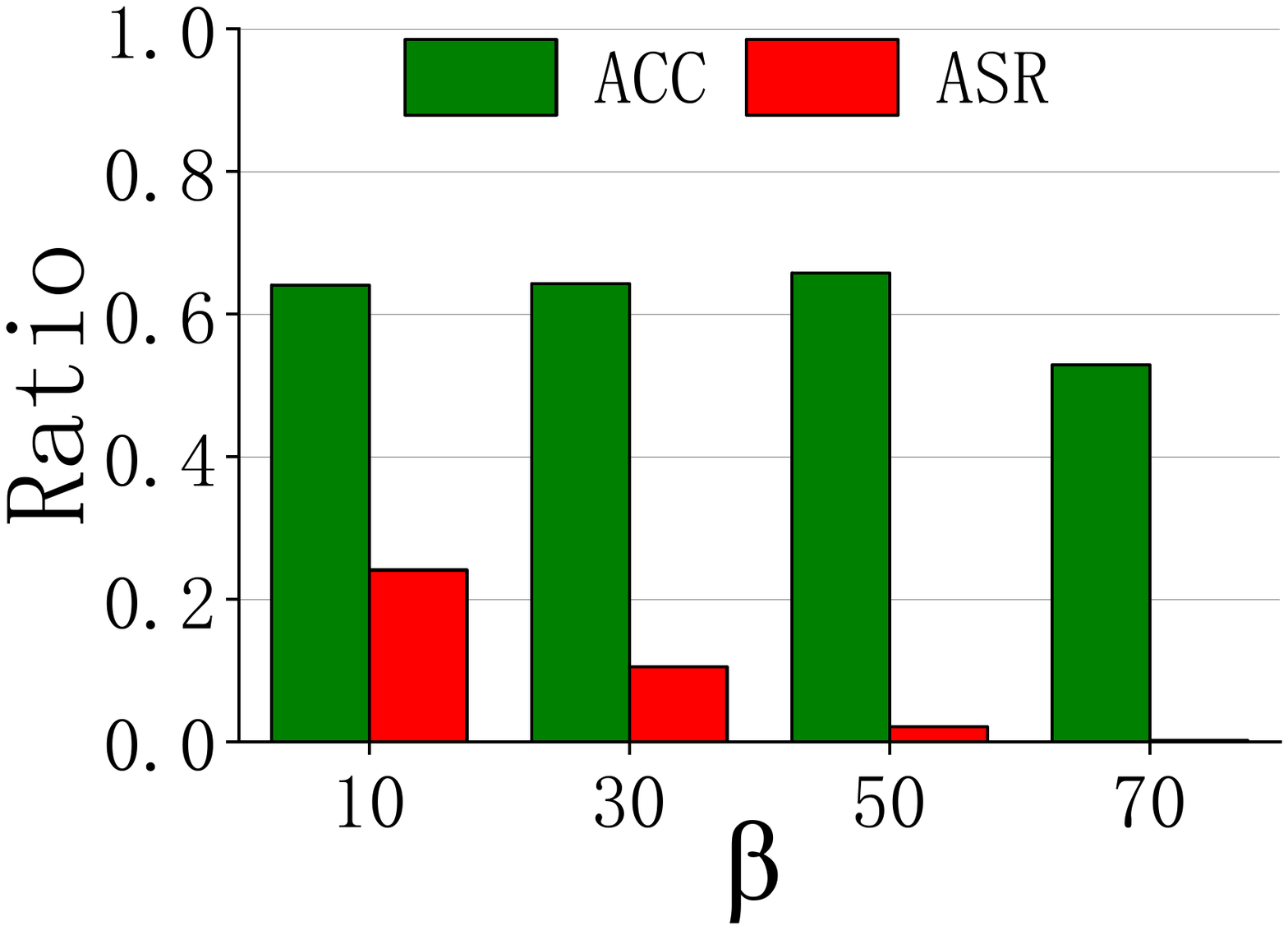}
  		\caption{WaNet}
    \end{subfigure}
\caption{Elimination performance with different $\beta$.}
\label{fig: beta}
\end{figure}

We evaluate the performance of the backdoor elimination part of our method with different $\beta$ on the Tiny ImageNet dataset and different attack methods. As shown in Figure~\ref{fig: beta}, the larger the value of $\beta$, the more thorough the elimination of the backdoor, but also the greater the impact on the accuracy of the model. For the example of the AdvDoor attack, it eliminates the backdoor in the model when we set $\beta$ as $70$, but the ACC decreases by up to $6.5\%$.
Therefore, this hyperparameter selection manner is to choose the largest value possible while ensuring the accuracy of the model, when the backdoor test set cannot be known in practical use.

\section{Conclusion}

This paper proposed Backdoor Consistency by analyzing the difference in middle features between normal and backdoor samples, which connected backdoor detection for DNN inputs and backdoor elimination for DNN models. We designed the BeniFul, including backdoor input detection and backdoor elimination.
For detection, we trained a VAE on the middle features of clean inputs and determined whether an input contains a backdoor trigger based on the VAE and model inference results. For elimination, we used feature distance loss to maximize the distance between the target model and the original backdoor model.
Extensive experimental results validated the effectiveness of our BeniFul. Considering that the middle features exist in any DNN model, our future work will extend to backdoor defense in other domains such as federated learning, reinforcement learning, and transfer learning.

\bibliography{beniful.bib}


\appendix

\section{Attack Performance with Different Dropout Ratio}\label{Appendix DR}
In an accidental experiment, we found that different dropout ratios (DR), in the model inference stage, differently affect the accuracy of clean and backdoor datasets. Then we train backdoor ResNet-34 models on the CIFAR-100 dataset. As shown in Table~\ref{tab: different dropout}, we randomly dropout $50\%$, $80\%$, $90\%$, and $95\%$ outputs from the fourth residual block of the ResNet-34, and test the model clean accuracy (ACC) and backdoor attack success ratio (ASR) under BadNets, AdvDoor, and WaNet backdoor attacks. We conclude that a larger dropout ratio will seriously affect the model performance, but will not significantly impact the effectiveness of the backdoor attack. To some extent, the experimental result indicates that the middle features are sparse for backdoor input.

\begin{table}[h]
\setlength\tabcolsep{9pt}
\centering
\begin{tabular}{ccccc}
\hline
\multirow{2}{*}{DR} & \multirow{2}{*}{ACC} & BadNets & AdvDoor & WaNet\\
 & & ASR & ASR & ASR \\
\hline
0\%	 & 0.795	&1.000	&0.999	&0.993\\
50\% & 0.776    &1.000	&0.999  &0.993\\
80\% & 0.656    &0.999  &0.999  &0.991\\
90\% & 0.553	&0.996	&0.986	&0.975\\
\textbf{95}\% & \textbf{0.334}	&\textbf{0.943}	&\textbf{0.925}	&\textbf{0.908}\\
\hline
\end{tabular}
\caption{Evaluation over Different Dropout Ratios.}
\label{tab: different dropout}
\vspace{-0.3cm}
\end{table}

\section{VAE Architecture Used in Experiments} \label{Appendix VAE architecture}
The VAE model in our experiment mainly uses three neural network layers: Convolutional Layer, Fully Connected Layer, and Transposed Convolutional Layer. We define a Convolutional Layer as $Conv_{k,s,p_1,p_2}^{c_1 \rightarrow c_2}$ and a Transposed Convolutional Layer as $ConvT_{k,s,p_1,p_2}^{c_1 \rightarrow c_2}$, where $k$, $s$, $p_1$, and $p_2$ represent the size of the convolution (or transposed convolutional) kernel as $k \times k$, convolution (or transposed convolutional) stride as $s \times s$, four sides of input padding size and output padding size, and $c_1$ and $c_2$ represent the number of input channels and output channels. We define a Fully Connected Layer as a $FC^{m \rightarrow n}$, where $m$ and $n$ represent the input and output dimensions. We use $ReLU$ as the nonlinear activation layer, $LayerNorm$ as the normalization layer, $MaxPool$ with $2 \times 2$ kernel size as the pooling layer, and $Upsample$ with twice the size magnification as the upsampling layer. And, $Sum$ represents adding the outputs of two neural networks together. We use Flatten Layer $FL$ to expand multi-dimensional features into one-dimensional features, and Unflatten is its inverse. The Reparameter Layer represents reparameterization option, $\mu + e^{\ln \sigma} \cdot \epsilon$, where $\mu$ is an output of $FC^{5670\rightarrow 64}_{mean}$, $\ln \sigma$  is an output of $FC^{5670\rightarrow 64}_{lnvar}$, and $\sigma$ is a vector with $64$ dimensions which elements correspond to the normal distribution. The specific organizational structure of these neural networks is shown in Table~\ref{tab: variational auto-encoder architecture}.

\begin{table}[h]
\setlength\tabcolsep{3pt}
\centering
\begin{tabular}{c|c|c|c}
\hline
layer name & \multicolumn{2}{c|}{layer}  & output size\\
\hline

-  & \multicolumn{2}{c|}{input} & $1 \times 256 \times 64$\\
\hline
\multirow{3}{*}{encoder1}  & \makecell[c]{$Conv_{3,1,1,0}^{1 \rightarrow 8}$ \\ $ReLu$ \\ $Conv_{3,2,1,0}^{8 \rightarrow 8}$ \\ $LayerNorm$} & \makecell[c]{$MaxPool$ \\ $Conv_{1,1,0,0}^{1 \rightarrow 8}$}  & \multirow{3}{*}{$8 \times 128 \times 32$}\\
\cline{2-3}
 &  \multicolumn{2}{c|}{\makecell[c]{$Sum$ \\ $ReLu$}}  &\\
\hline

\multirow{3}{*}{encoder2}  & \makecell[c]{$Conv_{3,1,1,0}^{8 \rightarrow 16}$ \\ $ReLu$ \\ $Conv_{3,2,1,0}^{16 \rightarrow 16}$ \\ $Layer Norm$} & \makecell[c]{$MaxPool$ \\ $Conv_{1,1,0,0}^{8 \rightarrow 16}$}  & \multirow{3}{*}{$16 \times 64 \times 16$}\\
\cline{2-3}
 &  \multicolumn{2}{c|}{\makecell[c]{$Sum$ \\ $ReLu$}}  &\\
\hline

\makecell[c]{encoder3} & \multicolumn{2}{c|}{\makecell[c]{$Conv_{5,2,0,0}^{16 \rightarrow 32}$ \\ $Layer Norm$ \\ $Flatten$}} & \makecell[c]{$5670$}  \\
\hline
\multirow{3}{*}{reparameter}  & \makecell[c]{$FC^{5670\rightarrow 64}_{mean}$} & \makecell[c]{$FC^{5670\rightarrow 64}_{lnvar}$} & \multirow{2}{*}{$64$}\\
\cline{2-3}
 & \multicolumn{2}{c|}{\makecell[cc]{$Reparameter$}} & \\
\cline{2-4}
  & \multicolumn{2}{c|}{$FC^{64\rightarrow 5670}$}& $5670$  \\
\hline

\makecell[c]{decoder1} & \multicolumn{2}{c|}{\makecell[c]{$Unflatten$ \\ $Layer Norm$ \\ $ConvT_{5,2,0,1}^{32\rightarrow 16}$ \\ $ReLu$}}  & \makecell[c]{$16 \times 64 \times 16$} \\
\hline

\multirow{3}{*}{decoder2} & \makecell[c]{$Layer Norm$ \\ $ConvT_{3,1,1,0}^{16\rightarrow 16}$ \\ $ReLu$ \\ $LayerNorm$ \\ $ConvT_{3,2,1,1}^{16\rightarrow 8}$} & \makecell[c]{$Upsample$ \\ $Conv_{1,1,0,0}^{16 \rightarrow 8}$} & \multirow{3}{*}{$8 \times 128 \times 32$} \\
\cline{2-3}
 & \multicolumn{2}{c|}{\makecell[c]{$Sum$ \\ $ReLu$}}  &   \\
\hline

\multirow{3}{*}{decoder3} & \makecell[c]{$Layer Norm$ \\ $ConvT_{3,1,1,0}^{8\rightarrow 8}$ \\ $ReLu$ \\ $LayerNorm$ \\ $ConvT_{3,2,1,1}^{8\rightarrow 1}$} & \makecell[c]{$Upsample$ \\ $Conv_{1,1,0,0}^{8 \rightarrow 1}$} & \multirow{3}{*}{$1 \times 256 \times 64$} \\
\cline{2-3}
 & \multicolumn{2}{c|}{\makecell[c]{$Sum$}}  &   \\
\hline
-  & \multicolumn{2}{c|}{output} & $1 \times 256 \times 64$\\
\hline
\end{tabular}
\caption{VAE Model Architecture.}
\label{tab: variational auto-encoder architecture}
\vspace{-0.3cm}
\end{table}

\section{Detailed Experimental Environments}
Our experimental computer is equipped with Intel(R) Xeon(R) Gold 6256 CPU @ 3.60GHz, NVIDIA GeForce RTX 4090 graphics card, and 160G memory. We adopt Ubuntu 20.04.6 LTS as the operating system and adopt CUDA 12.2 as the GPU computing framework. The software dependency packages for our code include PyTorch 2.2.2, scikit-learn 1.3.2, MLclf 0.2.14, numpy 1.24.3, pillow 10.2.0, and tqdm 4.61.2.

\section{Detailed Hyperparameters}
For the target DNN model, we use PyTorch to initialize the model's weights. To simulate the attacker, we train the model on poisoned training sets using an AdamW optimizer with a $0.00003$ learning rate. In BeniFul-BID, we use an AdamW optimizer with a $0.005$ learning rate to train the VAE model about $100$ epochs. In BeniFul-BE, we use an AdamW optimizer with a $0.00005$ learning rate to finetune the target model. We take 320 samples as one batch in all training processes.

\end{document}